\documentclass[preprint,superscriptaddress,prb]{revtex4}

\usepackage{graphicx}
\usepackage{epstopdf}
\usepackage{bbm}
\usepackage{amsmath}
\usepackage{eqnarray}

\usepackage{color}

\begin{document}

\title{High-field 1/f noise in hBN-encapsulated graphene transistors}

\author{A. Schmitt}\email{aurelien.schmitt@phys.ens.fr}
\affiliation{Laboratoire de Physique de l'Ecole normale sup\'erieure, ENS, Universit\'e
PSL, CNRS, Sorbonne Universit\'e, Universit\'e de Paris, 24 rue Lhomond, 75005 Paris, France}
\author{D. Mele}
\affiliation{Laboratoire de Physique de l'Ecole normale sup\'erieure, ENS, Universit\'e
PSL, CNRS, Sorbonne Universit\'e, Universit\'e de Paris, 24 rue Lhomond, 75005 Paris, France}
\affiliation{Univ. Lille, CNRS, Centrale Lille, Univ. Polytechnique Hauts-de-France, Junia-ISEN, UMR 8520-IEMN, F-59000 Lille, France.}
\author{M. Rosticher}
\affiliation{Laboratoire de Physique de l'Ecole normale sup\'erieure, ENS, Universit\'e
PSL, CNRS, Sorbonne Universit\'e, Universit\'e de Paris, 24 rue Lhomond, 75005 Paris, France}
\author{T. Taniguchi}
\affiliation{Advanced Materials Laboratory, National Institute for Materials Science, Tsukuba,
Ibaraki 305-0047,  Japan}
\author{K. Watanabe}
\affiliation{Advanced Materials Laboratory, National Institute for Materials Science, Tsukuba,
Ibaraki 305-0047, Japan}
\author{C. Maestre}\affiliation{Laboratoire des Multimat\'eriaux et Interfaces, UMR CNRS 5615, Univ Lyon, Universit\'e Claude Bernard Lyon 1,
F-69622 Villeurbanne, France}
\author{C. Journet}\affiliation{Laboratoire des Multimat\'eriaux et Interfaces, UMR CNRS 5615, Univ Lyon, Universit\'e Claude Bernard Lyon 1,
F-69622 Villeurbanne, France}
\author{V. Garnier}\affiliation{Universit\'e de Lyon, MATEIS, UMR CNRS 5510, INSA-Lyon, F-69621 Villeurbanne cedex, France}
\author{G. F\`eve}
\affiliation{Laboratoire de Physique de l'Ecole normale sup\'erieure, ENS, Universit\'e
PSL, CNRS, Sorbonne Universit\'e, Universit\'e de Paris, 24 rue Lhomond, 75005 Paris, France}
\author{J.M. Berroir}
\affiliation{Laboratoire de Physique de l'Ecole normale sup\'erieure, ENS, Universit\'e
PSL, CNRS, Sorbonne Universit\'e, Universit\'e de Paris, 24 rue Lhomond, 75005 Paris, France}
\author{C. Voisin}
\affiliation{Laboratoire de Physique de l'Ecole normale sup\'erieure, ENS, Universit\'e
PSL, CNRS, Sorbonne Universit\'e, Universit\'e de Paris, 24 rue Lhomond, 75005 Paris, France}
\author{B. Pla\c{c}ais} \email{bernard.placais@phys.ens.fr}
\affiliation{Laboratoire de Physique de l'Ecole normale sup\'erieure, ENS, Universit\'e
PSL, CNRS, Sorbonne Universit\'e, Universit\'e de Paris, 24 rue Lhomond, 75005 Paris, France}
\author{E. Baudin} \email{emmanuel.baudin@phys.ens.fr}
\affiliation{Laboratoire de Physique de l'Ecole normale sup\'erieure, ENS, Universit\'e
PSL, CNRS, Sorbonne Universit\'e, Universit\'e de Paris, 24 rue Lhomond, 75005 Paris, France}

\begin{abstract}
\textbf{Low-frequency 1/f noise in electronics is a conductance fluctuation, that has been expressed in terms of a mobility "$\alpha$-noise" by Hooge and Kleinpenning. Understanding this noise in graphene is a key towards high-performance electronics. Early investigations in diffusive graphene have pointed out a deviation from the standard Hooge formula, with a modified expression where the free-carrier density is substituted by a constant density $n_\Delta\sim10^{12}\;\mathrm{cm^{-2}}$. We investigate hBN-encapsulated graphene transistors where high mobility gives rise to the non-linear velocity-saturation regime. In this regime, the $\alpha$-noise is accounted for by substituting conductance by differential conductance $G$, resulting in a bell-shape dependence of flicker noise with bias voltage $V$.
The same analysis holds at larger bias in the Zener regime, with two main differences: the first one is a strong enhancement of the Hooge parameter reflecting the hundred-times larger coupling of interband excitations to the hyperbolic phonon-polariton (HPhP) modes of the mid-infrared Reststrahlen (RS) bands of hBN. The second is an exponential suppression of this coupling at large fields, which we attribute to decoherence effects. We also show that the HPhP bands
control the amplitude of flicker noise according to the graphene-hBN thermal coupling estimated with microwave noise thermometry. The phenomenology of $\alpha$-noise in graphene supports a quantum-coherent bremsstrahlung interpretation of flicker noise. }
\end{abstract}

\maketitle

%\section{Introduction}
Flicker (1/$f$) noise is a type of noise that dominates the spectral density at low frequencies. Ubiquitous in condensed matter and quantum conductors, it has also been measured in a large variety of systems ranging from biology to economics.
In electronics, flicker noise is a conductance fluctuation revealed as an excess low-frequency current noise, of spectral density $S_{I} = C I^{2}/f$, superimposed on the Johnson-Nyquist thermal noise $S_{th}=4G k_{B}T$ at finite bias current $I$.  Despite almost a century of research on $1/f$ noise since Johnson measured it in vacuum tubes [\onlinecite{Johnson1925pr}], its origin remains controversial. In semiconductors, $1/f$ noise is usually well described using the McWorther model relying on carrier number fluctuations due to defect traps [\onlinecite{McWorther,lowfreq}]. Alternative mechanisms based on carrier-number-fluctuation or involving fluctuations of mobility have been proposed.
As pointed out by Hooge and Kleinpenning [\onlinecite{Hooge1969pla}], the noise amplitude $C$ is a bulk effect that can be conveniently expressed in terms of a mobility noise, called $\alpha$-noise, with an intensive Hooge parameter $\alpha_{H}=NC $, where $N$ is the number of carriers participating to the conductance. Measurements in both metals [\onlinecite{Hooge1969physica}] and semiconductors [\onlinecite{Hooge1981rep}] have yielded a quasi-universal Hooge parameter $\alpha_{H} \sim 2-5 \times 10^{-3}$. A leading interpretation of the $\alpha$-noise relies on a distribution of two-level fluctuators that locally and elastically modulate electronic transmission [\onlinecite{Hooge1981rep},\onlinecite{Kogan}]. An alternative interpretation, based on a quantum theory proposed by Handel [\onlinecite{Handel1975,Handel1996pss}], points to an infrared (IR) bremsstrahlung origin which leads to a $1/f$ noise spectrum with a universal $\alpha_{H}=2\alpha_{0}/\pi = 4.6 \times 10^{-3}$ in the quantum-coherent regime, where $\alpha_{0}=1/137$ is the vacuum fine-structure constant that controls light-matter coupling. This interpretation puts emphasis on an inelastic origin of conductance noise; a critical discussion can be found in Refs.[\onlinecite{Nieuwenhuizen1987,Weissman1988}]. These contrasted interpretations take root in the ambivalent nature of conductivity, which is both  a momentum relaxation coefficient $\sigma_{1}(t)=(J/E)(t)$, and an energy relaxation parameter in the Joule power density $\sigma_{2}(t)=(P/E^{2})(t)$. Whereas $\sigma_{1}=\sigma_{2}$ in average, they may differ in their time dependencies due to different scattering mechanisms and times, fingerprinted in a low-frequency noise $\delta\sigma(t)$.

Graphene is an attractive platform for flicker noise investigation because of the tunability of charge carrier density and polarity. This versatile material also presents strong potential for electronics [\onlinecite{Balandin2013}] for which flicker noise is an important performance limit [\onlinecite{Mavredakis2013acsaem}]. Most experiments so far have been carried out at low to moderate bias with a $1/f$ corner frequency, separating the low-frequency $1/f$ noise from the high-frequency thermal white noise, in the sub-MHz range [\onlinecite{Lin2008,Rumyantsev2010}], and mostly on SiO$_2$-supported devices with a low mobility $\mu \lesssim 0.1\;\mathrm{m^{2}/Vs}$. Flicker noise in these diffusive graphene transistors was characterized by a doping-independent flicker amplitude $A=f S_{I}/I^{2} LW \in [10^{-7},10^{-6}]\;\mathrm{\mu m^{2}}$ [\onlinecite{Balandin2013}] (or $[0.1,1]\;\mathrm{nm^{2}}$), where $L$ and $W$ the length and width of the graphene channel. This doping-independent value for graphene violates the Hooge empirical formula where $A=\alpha_H/n$ is assumed to be inversely proportional to carrier density $n$, based on the assumption that electrons behave independently. It can be re-conciliated with a collective picture on substituting $n$ by a constant density  $n_\Delta$, which origin remains to be clarified. Some reports also point to a flicker noise dip at charge neutrality [\onlinecite{Rumyantsev2010,Takeshita2016}], but charge neutrality corresponds to a non-local mesoscopic regime that deviates from standard local-transport $\alpha$-noise conditions. The same occurs under intense  magnetic fields,  with a field-induced noise reduction [\onlinecite{Rehman2022nanoscale}], leading to a full suppression enforced by conductance quantization [\onlinecite{Kalmbach2016}]. Flicker noise suppression was also recently reported in the different system of a break-junction, where it is found to accompany shot-noise suppression at full transmission [\onlinecite{Shein2022}].
A deviation from the above $A=Cte$ diffusive graphene flicker noise was reported in moderate-mobility suspended samples, where a noise reduction was observed [\onlinecite{Kumar2015apl}]. 

The high-mobility hBN-encapsulated graphene transistors investigated in the present work constitute a testbed for $\alpha$-noise as they allow studying flicker noise in an extended electric-field range where different scattering and relaxation mechanisms succeed one another in increasing bias. In the perspective of testing the bremsstrahlung interpretation, the use of hBN-encapsulates is an additional asset, as hBN not only provides a superior mobility [\onlinecite{Dean2010nnano}], but also a very characteristic  mid-infrared (MIR) near-field electromagnetic environment [\onlinecite{Dai2015nnano,Kumar2015nl}], with its two Reststrahlen (RS) bands, $\hbar\Omega_I=95$-$100\;\mathrm{meV}$ and $\hbar\Omega_{II}=170$-$200\;\mathrm{meV}$. As a matter of fact, recent experiments have shown that the hyperbolic light of these RS bands strongly couples to graphene electronic transport [\onlinecite{Yang2018nnano,Baudin2020adfm}]. The three electronic transport regimes observed in increasing electric field are: (i) the mobility-limited Drude regime $J=\sigma E= ne \mu~ E$ at low field, (ii) the velocity saturation regime for $E\gtrsim E_{sat}=v_{sat}/\mu$ where $v_{sat}$ is the phonon-limited saturation velocity, and finally (iii) the interband Zener regime characterized by a doping and bias-independent differential conductivity $\sigma_Z$. Based on combined transport, flicker and thermal noise characterization in a series of high-quality samples, we show that : i) the standard analysis of flicker noise in graphene can be consistently extended to the non-linear saturation regime and, ii) the interband Zener contribution of gapless graphene can be accounted for by introducing a semi-empirical formula in the spirit of the bremsstrahlung interpretation, accounting for the enhanced electromagnetic coupling of interband excitations and decoherence effects.

\begin{figure}[h!]
\hspace*{-1cm}
\centering{}\includegraphics[width=14cm]{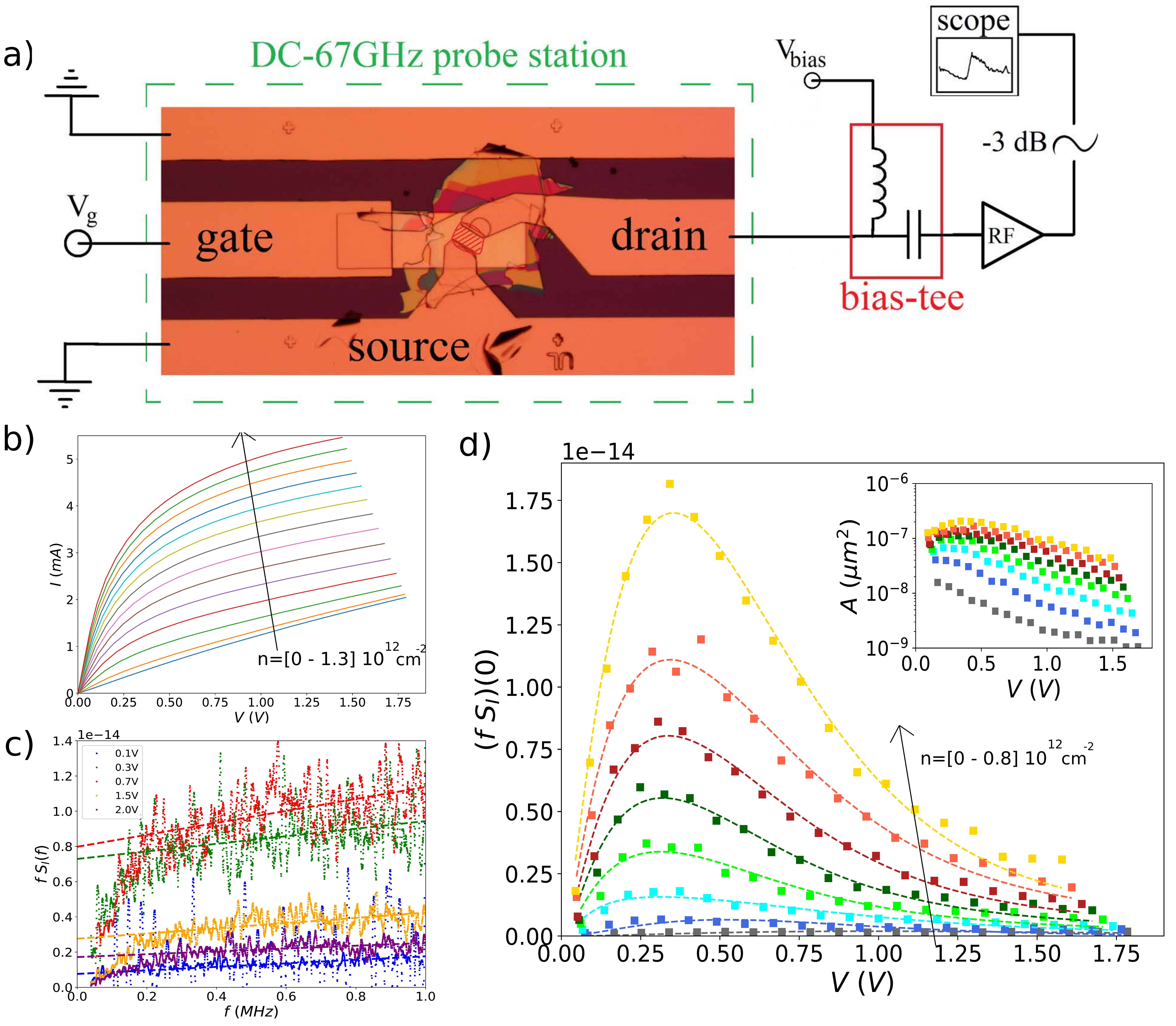}
\caption{DC transport and flicker noise in the high-mobility graphene transistor GRS5. a) Experimental setup for low-frequency and GHz noise measurements, with the graphene transistor embedded in a coplanar waveguide.  b): current-voltage curves for doping $n=[0-1.3]10^{12}\;\mathrm{cm^{-2}}$ corresponding to a Fermi energy $\epsilon_{F}=[0- 130]\;\mathrm{meV}$. c): Noise spectra in increasing bias at $n=0.6~ 10^{12} ~\mathrm{cm^{-2}}$. Flicker noise appears as plateaus of $fS_I(f)$. Dashed lines are linear fits of  $fS_I(f)$ of the data with slopes and extrapolates $(fS_{I})(0)$ measuring respectively the white noise and flicker noise. The fall down of noise below $0.2\;\mathrm{MHz}$ is due to the cut-off frequency of the bias-tee. d):  bell-shape dependence of $fS_{I}(0)$ versus $V$ for  $n=[0-0.8]10^{12}\;\mathrm{cm^{-2}}$; inset shows a semi-log plot of $A=f S_{I}/(GV)^{2} LW$ for same doping values indicating an exponential decay of $A$ with bias. }
\label{Fig1}
\end{figure}

We use a series of ten hBN-encapsulated graphene transistors, of a large size $(L,W)=4-25\;\mathrm{\mu m}$ and low edge-contact resistance $R_c$, exhibiting varied but large mobility $\mu=2$-$15\;\mathrm{m^2/Vs}$, saturation velocity $v_{sat}=0.2$-$0.8 v_F$, and Zener conductivity $\sigma_Z=0.1$-$1\;\mathrm{mS}$, as described in Supplementary Table-SI-1. They map a broad range of hBN dielectric thickness $t_{hBN}=50-160\;\mathrm{nm}$ and a variety of gating: (IR reflecting) Au bottom gates, (IR absorbing) graphite bottom gates or SiO$_2$-insulated Si back gates. We also use two different hBN grades: the high-pressure high-temperature (HPHT) from NIMS [\onlinecite{Taniguchi2007}], and the polymer-derived ceramic (PDC) from Lyon [\onlinecite{Maestre2022},\onlinecite{Pierret2022MatRes}] which behave differently. All devices are embedded in  coplanar waveguides for room-temperature probe-station measurement of DC transport, sub-MHz flicker noise, and microwave thermal noise (see Fig. \ref{Fig1}-a). Thanks to their high mobility,  transistors sustain large saturation currents ($\sim 1\;\mathrm{mA/\mu m}$), rejecting the $1/f$ noise corner frequency ($f_c=CI^2/4Gk_BT\propto I$) in the low GHz range. As a matter of fact, the flicker noise tails are visible in the GHz noise spectra and allow for a sanity check of the sub-MHz measurements.

Current noise is analyzed with a modified Hooge formula $S_I=C(GV)^2/f$, obtained by substituting the total DC current $I$ by the differential current $GV$ to account for  non-linear transport [\onlinecite{Mavredakis2020acsaem}]. Here $V$ is the drain-source voltage obtained after subtraction of the contact-voltage drop ($R_cI$), $G=(ne\mu(E)+\sigma_Z)W/L$ is the  differential conductance with $\mu(E)=\mu(0)/(1+E/E_{sat})^2$ the optical phonon-limited mobility with $E_{sat}$ and $v_{sat}=\mu(0) E_{sat}$ the saturation field and velocity. Following Ref.[\onlinecite{Yang2018nnano}], we correct for drain-gating effect by biasing transistors along constant density lines on applying a bias-dependent gate voltage $V_{g}(V)=V_g(0)+ \beta V$. This biasing procedure allows rejecting drain pinch-off effect, such as investigated in Ref.[\onlinecite{Schmitt2022}], while securing quasi-homogeneous doping and electric fields, up to small gradients that become negligible at large doping. 
Low-frequency noise characterization is enriched by Johnson-Nyquist thermal noise measurements, performed in the microwave ($1$-$10\;\mathrm{GHz}$) band where flicker noise contribution is negligible [\onlinecite{Yang2018nnano}]. This allows estimating the electronic temperature $T_N$ and extract the  thermal conductivity $dP/dT_N$ to the hBN substrate which turns on in the Zener regime [\onlinecite{Baudin2020adfm}], where $P=VI/LW$ is the areal Joule power.

%\section{Sample description and measurements}

Figure \ref{Fig1} summarizes the experimental technique (panel a) and procedure (panels b-d). Figure \ref{Fig1}-b shows typical current-voltage curves measured in the typical transistor ``GrS5". They show the successive low-bias linear ohmic behavior, the velocity-saturation regime  for $V > V_{sat}\simeq0.25\;\mathrm{V}$, and finally the interband Zener regime. The saturation of intraband current can be consistently attributed to the scattering by the optical phonons of the lower Reststrahlen  (RS) band of hBN ($\hbar \Omega_{I}= 95-100\;\mathrm{meV}$) [\onlinecite{Yang2018nnano}]. The Zener regime is accompanied by the onset of energy relaxation by optical phonons of the upper RS band of hBN ($\hbar \Omega_{II}= 170-200\;\mathrm{meV}$) [\onlinecite{Yang2018nnano}]. Its  main signature is electron cooling, that is observed in the noise temperature (inset of Fig.\ref{Fig3}-c) by a decrease of the temperature-field slope. This change of slope is typical of single-layer graphene; in bilayer graphene this radiative cooling is even more drastic and eventually takes the form of a temperature drop with bias [\onlinecite{Baudin2020adfm}].

The differential conductivity (not shown) is accurately fitted using the sum of saturation ($\sigma_{sat}$) and Zener ($\sigma_{Z}$) contributions [\onlinecite{Meric2008,Yang2018nnano}] :
\begin{equation}
\label{conductance}
    \sigma(E)=ne \frac{\mu(0)}{(1+E/E_{sat})^{2}}+ \sigma_{Z}\qquad,
\end{equation}
 where $\mu(0)$ is the extrapolated low-bias electronic mobility, $n$ is the carrier density,  and $E=V/L$. $\mu(0)$  agrees with standard mobility $\mu$ deduced from the low-bias $G(V_g)$ dependence. $G=\sigma W/L$, and its constituents $G_{sat}$ (or $\sigma_{sat}$) and $G_Z$ (or $\sigma_{Z}$), are the associated differential conductance (conductivity). Eq.(\ref{conductance}) is used to extract DC transport parameters $\mu(0)$, $E_{sat}$ or $v_{sat}=\mu(0) E_{sat}$ that are listed in Table-SI1.

For each bias and doping, we measure the current noise in the $f=[0.1-1]\;\mathrm{MHz}$ frequency band, as shown in Fig.\ref{Fig1}-c. Flicker $1/f$ noise corresponds to plateaus $fS_{I}(f)=a+bf$, where $a$ is the amplitude of flicker noise and $b$ measures the instrumental background noise. The $\alpha$-noise amplitude $a$ (denoted $fS_I(0)$ in Fig.\ref{Fig1}-d and thereafter) has an asymmetric bell-shape bias dependence, and a strong doping dependence $fS_{I}\propto n^{2}$. The noise amplitude  $A=fS_{I}/(GV)^{2}(WL)$ is plotted in the inset of Fig.\ref{Fig1}-d. Its low-bias extrapolate $A\sim 10^{-7}\;\mathrm{\mu m^{2}}$ is consistent with diffusive graphene values [\onlinecite{Balandin2013}]. At high bias we observe an amplitude suppression,  with  $A$ dropping below $10^{-8}\;\mathrm{\mu m^{2}}$, following an exponential dependence $A\propto e^{-E/E_{\Lambda}}$ (inset of Fig.\ref{Fig1}-d), with a characteristic electric field $E_{\Lambda}\sim 0.1\;\mathrm{V/\mu m}$. The doping dependence  $fS_{I} \propto n^{2}$ is quite generic; it suggests a doping-independent velocity flicker noise, $fS_{v}LW=A v^2$ (with $v=\mu(E)E$), as illustrated in Supplementary Information Fig.SI-2.

\begin{figure}[h!]
\hspace*{-1cm}
\centering{\includegraphics[width=17cm]{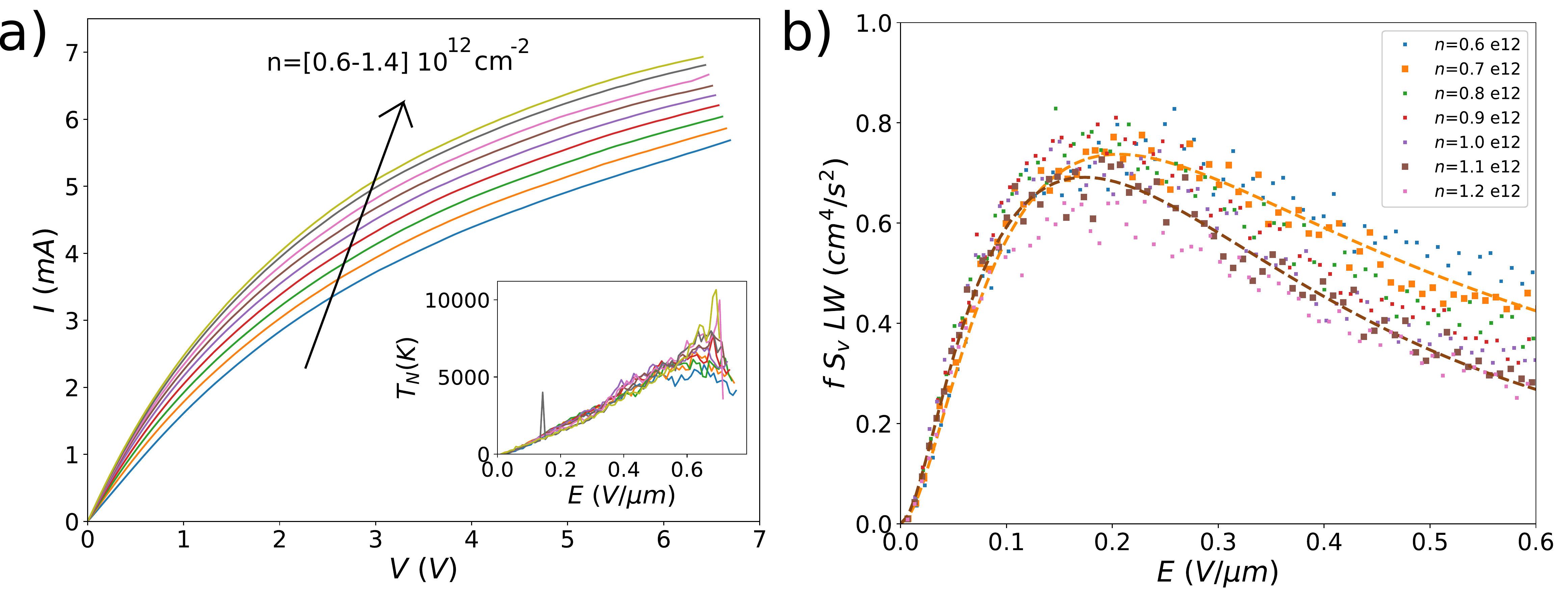}} \caption{DC transport and flicker noise in the high-mobility graphene transistor ``Lyon1".
a): current-voltage curves, for doping $n=[0.6-1.4]10^{12}\;\mathrm{cm^{-2}}$, and noise thermometry (inset) show no fingerprint of Zener tunneling and cooling.
b): Quasi-scaling of the velocity flicker noise $fS_vLW(E)$ as a function of electric field $E=V/L$. The representative $n=0.7\;10^{12}\;\mathrm{cm^{-2}}$ (resp.  $n=1.1\;10^{12}\;\mathrm{cm^{-2}}$) data are well fitted by
Eq.(\ref{saturation-flicker}) with $A=6.4\;10^{-7}\;\mathrm{\mu m^2}$ and $v_{sat}=0.82~v_{F}$ (resp. $A=4.6\;10^{-7}\;\mathrm{\mu m^2}$ and $v_{sat}=0.65~v_{F}$).}
\label{Fig2}
\end{figure}

We start our analysis by focusing on the velocity saturation regime, which is exemplified in sample ``Lyon1" where the Zener contribution to transport is minimized to $\sigma_Z\simeq0.1\;\mathrm{mS}$ (Table SI-1), as deduced from the fitting of  the differential conductance with Eq.(\ref{conductance}) in Fig.\ref{Fig2}-a data. 
Consistently, noise thermometry (inset of Fig.\ref{Fig2}-a) shows little fingerprint of Zener cooling, as opposed to other samples where Zener cooling shows up as a prominent breakout of the $T_N(E)$ dependence (see Supplementary Informations Fig.SI-3). The origin of Zener transport suppression in sample ``Lyon1" is discussed below. Importantly, the flicker noise in Fig.\ref{Fig2}-b reduces to its velocity-saturation contribution $fS_{sat}LW=A [\mu(E)E]^2$, which is entirely determined by the differential mobility $\mu(E)$, according to
\begin{equation}
\label{saturation-flicker}
fS_{sat}LW=A \times v_{sat}^2  \left[\frac{E/E_{sat}}{(1+E/E_{sat})^{2}} \right]^{2} \quad .
\end{equation}
Velocity-flicker in Eq.(\ref{saturation-flicker})  has a quadratic low-bias onset $fS_{sat}LW=A(\mu(0) E)^2$, consistent with standard Hooge law $fS_{I}LW=A(I)^2$, a peak $fS_{sat}LW=Av_{sat}^2/16$ at $E=E_{sat}$, and a bias tail $fS_{sat}LW=Av_{sat}^2(E_{sat}/E)^2$. Orange and brown dashed lines in Fig.\ref{Fig2}-b are theoretical fits with Eq.(\ref{saturation-flicker}) of the  $n=0.7$ and $n=1.1\;10^{12}\;\mathrm{cm^{-2}}$ data. The quasi-scaling observed here (and in Figs.SI-2 for the other samples) relies on the weak doping-dependence of $v_{sat}$ (and $E_{sat}= v_{sat}/\mu(0)$ as $\mu(0)$ is doping independent). From the amplitude of the $n=7.10^{11}\;\mathrm{cm^{-2}}$ data  we deduce  $A\simeq 6.4 \;10^{-7}\;\mathrm{\mu m^2}$ which is indeed consistent with diffusive-graphene literature data [\onlinecite{Balandin2013}]. Setting $A=2\alpha_{0}/\pi n_\Delta$ according to the modified Hooge-Handel formula, we infer $n_\Delta\simeq 0.7\; 10^{12}\;\mathrm{cm^{-2}}$.

\begin{figure}[h!]
\hspace*{-1cm}
\centering{}\includegraphics[width=18cm]{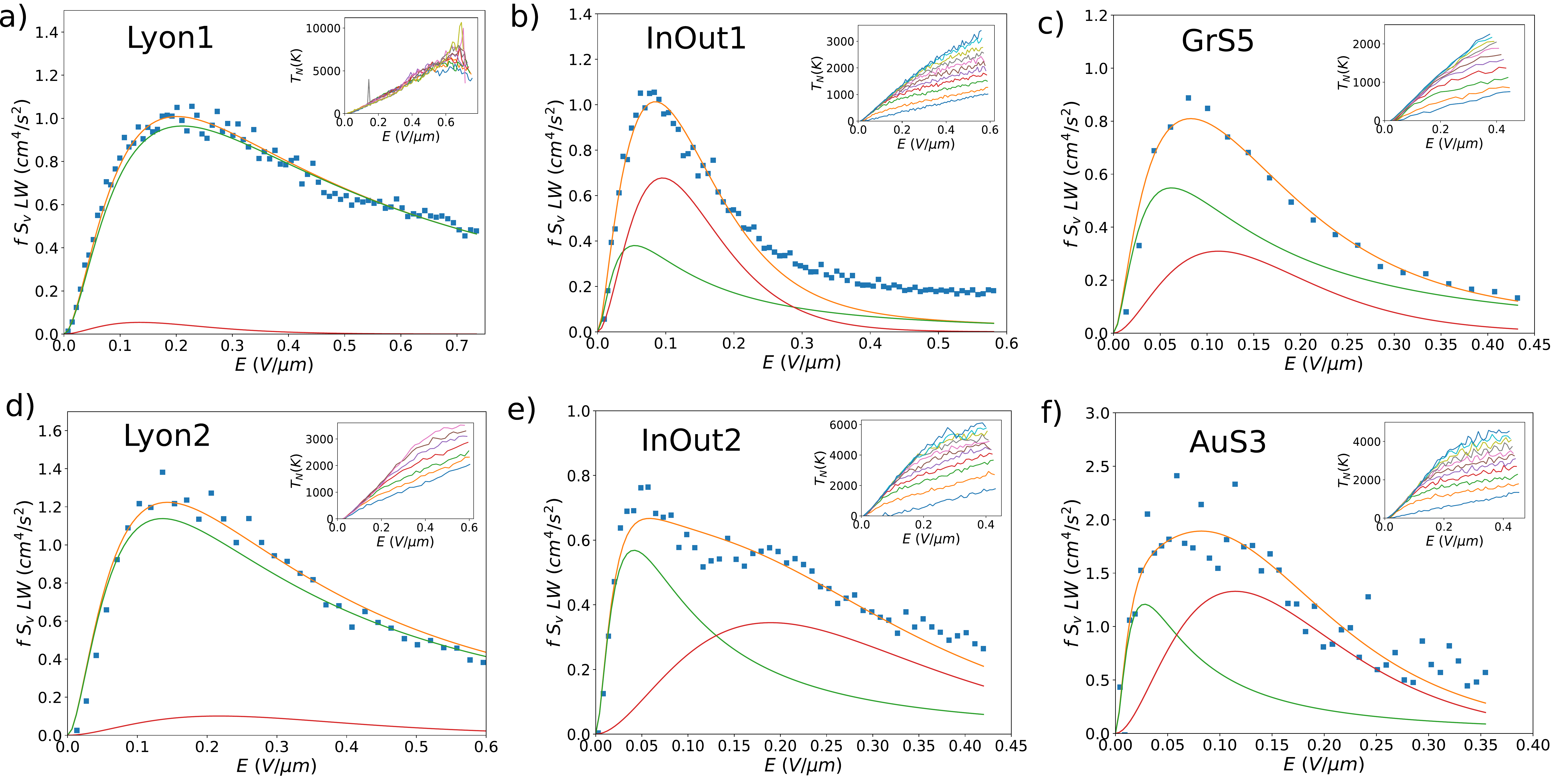} \caption{Velocity flicker noise $fS_{v}LW$ for $6$ typical devices at a representative doping $n=6\;10^{11}\;\mathrm{cm^{-2}}$ (dots), and their fits with Eqs.(\ref{saturation-flicker}) and (\ref{Zener-flicker}) for the two-fluid model. Green and red lines correspond to the saturation and Zener contributions respectively, and the orange lines correspond to their sum. Successive panels correspond to devices of increasing mobility $\mu=4\rightarrow15~ \mathrm{m}^{2}/\mathrm{Vs}$ (see Table SI-1). Values of the fitted parameters $A$ and $E_\Lambda$ are summarized in Table SI-1. Insets show for each device the noise temperature $T_{N}$ as function of bias and doping measured in the $f=[1-10]\;\mathrm{GHz}$ range, as detailed in the Supplementary Information section (Fig.SI-3)}
\label{Fig3}
\end{figure}

We now turn to the more general case where a significant Zener 
contribution is observed at high field, in both the DC transport (Fig.SI-1) and in the noise thermometry (Fig.SI-3 and insets of Figs.\ref{Fig3}),  and quantified in Table-SI1.
Figs.\ref{Fig3} compare velocity flicker noise at the representative doping $n=6.10^{11}\;\mathrm{cm^{-2}}$ for $6$ devices of the series. As seen in the figure, the bias dependencies generally deviate from the velocity-saturation bell-shape of Fig.\ref{Fig2}-b (reproduced in Fig.\ref{Fig3}-a) and described by Eq.(\ref{saturation-flicker}). To explain this phenomenology we include in the velocity flicker a semi-empirical Zener correction to the current flicker formula in the form
\begin{equation}
\label{Zener-flicker}
fS_{Zener}LW= B e^{-E/E_{\Lambda}}\times v_{Z}^2\left[\frac{E}{E_{\Lambda}}\right]^{2}  \quad ,
\end{equation}
where $B=\alpha_g/n_\Delta\sim100\times A$, with $\alpha_g=e^2/4\pi\epsilon_0\epsilon_{hBN}\hbar v_F\simeq0.70$ the fine-structure constant for hBN-encapsulated graphene ($\epsilon_{hBN}=3.4$ [\onlinecite{Pierret2022MatRes}]), $E_{\Lambda}$ the decoherence field discussed in Fig.\ref{Fig1}-d (inset), and $v_{Z}=\frac{\sigma_{Z}}{en_{\Delta}} E_{\Lambda} \sim 0.01 v_{F}$ a characteristic Zener velocity deduced from Zener conductivity $\sigma_{Z}$,  the characteristic doping $n_{\Delta}$, and $E_{\Lambda}$. The Zener contribution also has a bell-shape bias dependence, which is however different with a $E^2$ low bias dependence, a peak  $fS_{Zener}LW=4Bv_{Z}^2/e^2$ at $E=2E_{\Lambda}$ and an exponential tail. Inspection of Eqs.(\ref{saturation-flicker}) and (\ref{Zener-flicker}) shows that the two contributions have similar peak amplitudes for $v_{sat}/v_Z\sim8e^{-1}\sqrt{\frac{\pi\alpha_{g}}{2\alpha_{0}}}\simeq36$. Their relative amplitudes depend on $n_{\Delta}$ and $E_{\Lambda}$, which enter in $v_{Z}$ ($A$ and $B$ are simply proportional), and constitute the $2$ fitting parameters used below. Their peak positions are shifted by a factor $2$ for $E_{sat}\sim E_{\Lambda}$, in such a way that $S_{Zener}$ prominently affects the high-field tail of the flicker noise, when $S_{sat}$ accounts for the flicker noise onset.

 Figs.\ref{Fig3} also show fits of the velocity flicker data (orange lines) with a two-fluid model where $S_{v}=S_{sat}+S_{Zener}$ from Eqs.(\ref{saturation-flicker}) and (\ref{Zener-flicker}), using  $n_\Delta$ and $E_\Lambda$ as the only free parameters. The velocity-saturation  (green lines) and Zener (red lines) components are displayed in each panel. A good agreement with the two-fluid model is observed with, however, some deviations at extremely high bias $E\gtrsim0.5\;\mathrm{V/\mu m}$ in few devices (see e.g. "InOut1" in Fig.\ref{Fig3}-b), with an high-bias noise tail at a non-zero value, possibly due to an additional spurious mechanism that remains to be clarified. The Zener contribution is tiny in the "Lyon1" device (Fig.\ref{Fig3}-a), justifying the analysis of Fig.\ref{Fig2}-b in terms of the only saturation contribution. A similar trend is observed in "Lyon2"  (Fig.\ref{Fig3}-d) which is made with the same PDC-grade [\onlinecite{Maestre2022}] hBN material. Other devices, fabricated with high-pressure high-temperature HPHT-grade hBN from NIMS [\onlinecite{Taniguchi2007}], exhibit more prominent HPhP cooling in the Zener regime (see  Figs.\ref{Fig3} insets and Figs.SI-3). In these devices, single-fluid fits are unable to map the observed field dependencies, hence the justification of our 'two-fluid' (intraband saturation and interband Zener) analysis. The saturation contribution (green lines in Figs.\ref{Fig3}) is mostly responsible for the onset of flicker noise at low bias, with a slope increasing with mobility $\mu(0)$. The Zener contribution (red lines in Figs.\ref{Fig3}) becomes significant at larger bias, with a characteristic peak field at $2E_{\Lambda} > E_{sat}$, explaining the slower decrease of flicker noise at high bias. The amplitudes of the two contributions are comparable with, as a general trend, an intraband saturation term becoming more prominent in  high-mobility devices.

\begin{figure}[h!]
\hspace*{-1cm}
\centering{}\includegraphics[width=18cm]{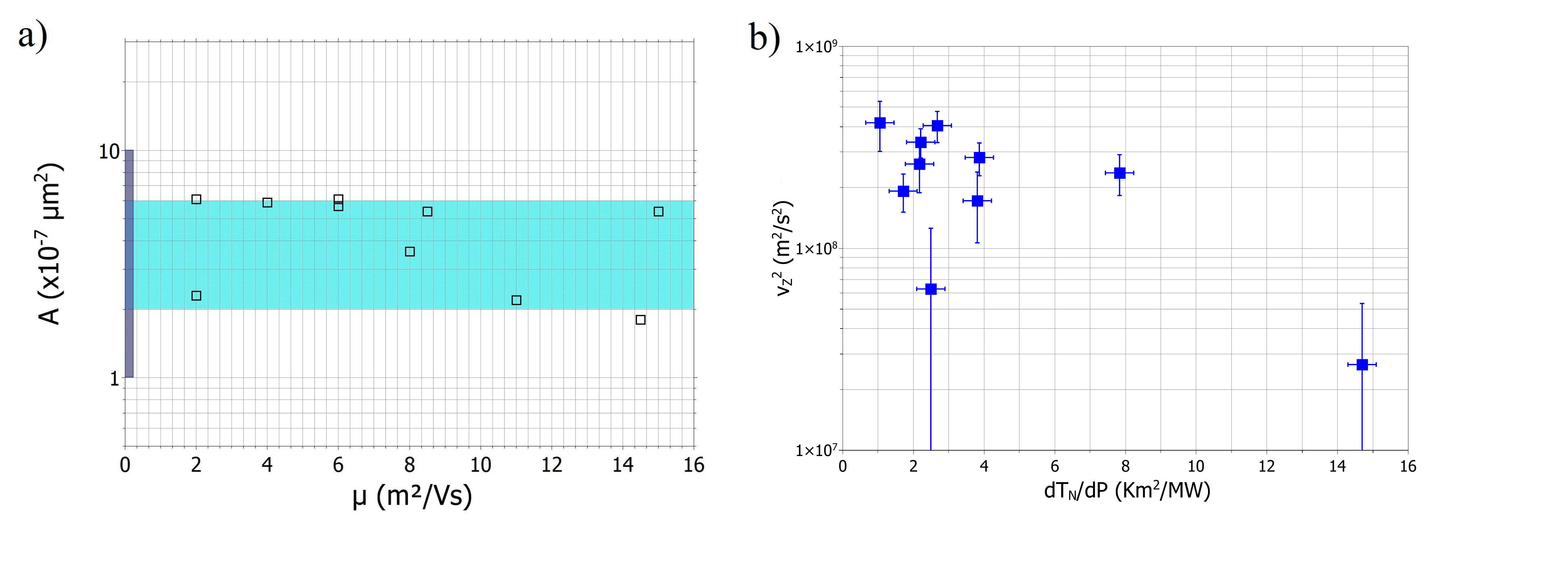} \caption{Flicker-noise  correlations to mobility and Zener cooling. Panel a): Noise amplitude $A$ as function of mobility for 10 devices of the series (dots). The light-blue rectangle corresponds to our hBN-encapsulated graphene devices. Panel b): Zener-flicker amplitude $v_Z^2$ as function of the thermal resistance $dT_N/dP$ in the Zener interband regime extracted from GHz noise thermometry measurements (insets of Fig.\ref{Fig3}) for the device series.}
\label{Fig4}
\end{figure}

We now discuss the sample-dependent parameters $A$ (or $n_{\Delta}$) and $E_{\Lambda}$, which are extracted from above fits and listed in Table-SI1. We obtain  $A\in[2-7]10^{-7}\;\mathrm{\mu m^2}$, corresponding to $n_{\Delta} \in[0.7-2.3]10^{12} \;\mathrm{cm^{-2}}$. These values, obtained in a broad $\mu(0)\in[2-15]\;\mathrm{ m^{2}/Vs}$ mobility range, confirm the literature data $A\in [10^{-7}-10^{-6}]\;\mathrm{\mu m^2}$ reported for diffusive  SiO$_2$-supported graphene (dark-blue rectangle in Fig. \ref{Fig4}-a), suggesting that $A$ is essentially impurity-scattering independent. This observation is further illustrated in the comparison between devices ``Goyave2" (Table SI-1) and ``AuS3" (Fig.\ref{Fig3}-f), which yield the same $A$ in spite of a factor $7$ in mobility. The similarity of flicker noise amplitude $A$ in the impurity-dominated (linear low-mobility) and OP-phonon-dominated (non-linear high-mobility) scattering regimes, reflects one more time its universality. The characteristic field for the Zener correcting term, $E_{\Lambda} = 75 \pm 25 ~\mathrm{mV/\mu m}$, is quite similar for the $6$ representative devices. It can be translated in terms of a Zener junction length $\Lambda = \sqrt{\hbar v_{F}/e E_{\Lambda}} = 100 \pm 15 ~\mathrm{nm}$, which shows a reduced dispersion when compared with the quite spread values of the Zener conductivity $\sigma_{Z}\in[0.1-1]\;\mathrm{mS}$. 
%The above statistics can be extended by including the full $10$ devices series, with similar results albeit with a somewhat larger scatter (see Table-SI1). 

As seen in Fig.\ref{Fig4}-a all values of $A$ lie in the light-blue domain $A\in[2-6]10^{-7}\;\mathrm{\mu m^{2}}$, where the two bounds correspond to $n_\Delta\in[0.7-2.3]10^{12}\;\mathrm{cm^{-2}}$. We can convert this doping range in terms of an energy range $\tilde{\Delta}=\hbar v_F\sqrt{\pi n_\Delta}$ relying on the massless dispersion of graphene.  Doing so we find that  $\tilde{\Delta}\in[\Delta_1,\Delta_2]=[0.09-0.17]\;\mathrm{eV}$ is actually bounded by the lower and upper Reststrahlen bands of hBN. These two energies have been associated in Ref.[\onlinecite{Yang2018nnano}] to momentum relaxation (out-of-plane optical phonons of lower RS band) and energy relaxation (in-plane optical phonons of upper RS band) respectively. This finding suggests that flicker noise in high-mobility graphene does map the ambivalent nature of conductance as an admixture of momentum and energy relaxations parameters. 

%\section{Interpretation}
%Let us first discuss the intraband flicker term. In hBN-encapsulated graphene, it is described by a universal $A=(4 \pm 2) 10^{-7}\;\mathrm{\mu m^{2}}$ that is consistent with diffusive-graphene literature. The values of $A$ for the 10 devices of the series are reported as a function of mobility in Figure \ref{Fig4}-a, where the dark-blue rectangle reminds the experimental window for low-mobility graphene, and where the mobility independence of $A$ is depicted.

The correlation of flicker noise to energy relaxation is even more obvious in the Zener contribution characterized by a large $B\approx 100\;A$ and the characteristic velocity $v_Z=\sigma_ZE_\Lambda/en_\Delta$ in Eq.(\ref{Zener-flicker}). The semi-log plot $v_Z^2(dT_N/dP)$ in Fig.\ref{Fig4}-b indeed shows a strong decrease of the Zener flicker amplitude with the thermal resistance to the hBN substrate.   This is exemplified by transistor "Lyon1", with a large thermal resistance to the hBN substrate due to weak Zener cooling, and a tiny Zener flicker amplitude. We attribute this reduced Zener cooling  to the larger structural disorder in the two "Lyon"-grade hBN-based devices, where hBN is fabricated through the PDC route.  As a matter of fact, hBN structural quality is essential to ensure free propagation of hyperbolic phonon polaritons in the hBN bulk; their strong backscattering/damping in "Lyon"-grade hBN devices is likely to suppress the MIR electromagnetic coupling, implying lower Zener conductivity and radiative cooling, resulting in  a smaller Zener flicker noise.

We now interpret the interband Zener term in Eq.(\ref{Zener-flicker}), which is introduced as a correction to the intraband saturation term Eq.(\ref{saturation-flicker}). It  is characterized by a large coupling constant $\alpha_{g} \sim 100 \times\alpha_{0}$, that is affected by a field reduction factor $e^{-E/E_{\Lambda}}$ with a nearly sample-independent characteristic field $E_\Lambda\simeq75\;\mathrm{mV/\mu m}$.
 This characteristic field  can be associated to the energy scale $\varepsilon_\Lambda=\sqrt{e\hbar v_F E_\Lambda }=7\;\mathrm{meV}$, and length $\Lambda=\hbar v_F/\varepsilon_\Lambda=100\;\mathrm{nm}$ of Zener tunneling. In this interpretation, $\varepsilon_\Lambda$ and $\Lambda$ are the characteristic energy and length of coherent tunneling, which are limited by inelastic HPhP scattering. They control the Hooge parameter $\alpha_{g}e^{-E/E_{\Lambda}}$ in its quantum-coherent version by Handel. Note in passing that the incoherent variant of quantum theory of flicker noise involves a large reduction factor $v_F^2/c^2\ll1$, where $c$ is the speed of light, with respect to the coherent one  $\alpha_{H}=2\alpha_0/\pi$ [\onlinecite{Handel1996pss}]. Further insight into this coherent mechanism requires a comprehensive theoretical approach, including geometrical antenna effects in free-light and HPhP emissions, which is still elusive and beyond the scope of the present experimental report.

%\section{Conclusion}
The conclusion of our experimental report of $\alpha$-noise in hBN-encapsulated graphene is twofold. In a first part we recover and generalize the Hooge formula for non-linear intraband transport by substituting the total current $I$ with the differential current $GV$. We give an interpretation of the constant flicker noise prefactor, previously revealed in diffusive graphene, in terms of a scattering density $n_\Delta$, defined by an effective energy $\tilde{\Delta}$ interpolating between the elastic-scattering energy $\Delta_{1}$  and the inelastic-scattering energy $\Delta_{2}$. In a second part, we extend the Hooge formula to include the interband transport contribution, which is prominent in high-mobility graphene at high field. We propose a compact formula, in the spirit of the Hooge-Handel interpretation of flicker noise, with a strongly enhanced interband Hooge parameter reflecting the large coupling of graphene electron-hole pairs to near-field hyperbolic phonon polaritons of hBN [\onlinecite{Baudin2020adfm}]. The graphene case  corresponds to the collective strongly-inelastic variant, where IR bremsstrahlung constitutes the main energy-relaxation mechanism in the Zener regime where $\alpha_H\sim1$. The superposition of the two flicker terms in the two-fluid $\alpha$-noise analysis accounts for the observed field dependencies in a series of graphene transistors with varied IR environments. Experiment indicates that the interband coupling constant is  affected by an exponential suppression factor accounting for decoherence effects in the HPhP coupling. Our observations that i) intraband Hooge parameter approaches the Handel limit $\alpha_{H,intra}=2\alpha_0/\pi$ , ii) interband parameter $\alpha_{H,inter}\sim\alpha_g$ is strongly enhanced by a factor $100$, and iii) that it is sensitive to decoherence effects via the $\exp(-E/E_\Lambda)$ factor, iv) and the correlation of flicker noise  with thermal conductance to the hBN substrate, constitute four qualitative indications supporting the quantum-coherent bremsstrahlung interpretation of flicker noise.
In addition, this electromagnetic coupling interpretation  provides a natural explanation for the ubiquitous scatter in flicker data by  ascribing it to the  variability in the IR environment. 
In a broader perspective, our work promotes flicker noise, in complement to DC transport and noise thermometry,  as a powerful semi-quantitative tool to decipher the physical mechanisms at stake in graphene transistors under extreme bias.

\section{Data availability}

Data are available at the DOI : https://doi.org/10.5281/zenodo.7632159

\section{Author declaration}
\subsection{Conflict of interest}
The authors have no conflict of interest to disclose.
\subsection{Author contributions}
AS, EB and BP conceived the experiment. AS conducted device fabrication and measurements, under the guidance of DM and MR in the early developments. TT,  KW, CM, CJ and VG have provided the hBN crystals. AS, EB and BP developed the models and theoretical interpretations. AS, DM, CV, JMB, GF, CV, EB and BP participated to the data analysis. AS wrote the manuscript with assistance of EB and BP, and contributions from the coauthors.

\subsection{Acknowledgments}

Authors thank Gerbold M\'enard for his critical reading of the manuscript. The research leading to these results has received partial funding from the European Union Horizon 2020 research and innovation program under grant agreement No.881603 "Graphene Core 3", and from the French  ANR-21-CE24-0025-01 "ELuSeM".

\end{document}

\title{High-field 1/f noise in hBN-encapsulated graphene transistors : Supplementary Information}

\author{A. Schmitt}\email{aurelien.schmitt@phys.ens.fr}
\affiliation{Laboratoire de Physique de l'Ecole normale sup\'erieure, ENS, Universit\'e
PSL, CNRS, Sorbonne Universit\'e, Universit\'e de Paris, 24 rue Lhomond, 75005 Paris, France}
\author{D. Mele}
\affiliation{Laboratoire de Physique de l'Ecole normale sup\'erieure, ENS, Universit\'e
PSL, CNRS, Sorbonne Universit\'e, Universit\'e de Paris, 24 rue Lhomond, 75005 Paris, France}
\affiliation{Univ. Lille, CNRS, Centrale Lille, Univ. Polytechnique Hauts-de-France, Junia-ISEN, UMR 8520-IEMN, F-59000 Lille, France.}
\author{M. Rosticher}
\affiliation{Laboratoire de Physique de l'Ecole normale sup\'erieure, ENS, Universit\'e
PSL, CNRS, Sorbonne Universit\'e, Universit\'e de Paris, 24 rue Lhomond, 75005 Paris, France}
\author{T. Taniguchi}
\affiliation{Advanced Materials Laboratory, National Institute for Materials Science, Tsukuba,
Ibaraki 305-0047,  Japan}
\author{K. Watanabe}
\affiliation{Advanced Materials Laboratory, National Institute for Materials Science, Tsukuba,
Ibaraki 305-0047, Japan}
\author{C. Maestre}\affiliation{Laboratoire des Multimat\'eriaux et Interfaces, UMR CNRS 5615, Univ Lyon, Universit\'e Claude Bernard Lyon 1,
F-69622 Villeurbanne, France}
\author{C. Journet}\affiliation{Laboratoire des Multimat\'eriaux et Interfaces, UMR CNRS 5615, Univ Lyon, Universit\'e Claude Bernard Lyon 1,
F-69622 Villeurbanne, France}
\author{V. Garnier}\affiliation{Universit\'e de Lyon, MATEIS, UMR CNRS 5510, INSA-Lyon, F-69621 Villeurbanne cedex, France}
\author{G. F\`eve}
\affiliation{Laboratoire de Physique de l'Ecole normale sup\'erieure, ENS, Universit\'e
PSL, CNRS, Sorbonne Universit\'e, Universit\'e de Paris, 24 rue Lhomond, 75005 Paris, France}
\author{J.M. Berroir}
\affiliation{Laboratoire de Physique de l'Ecole normale sup\'erieure, ENS, Universit\'e
PSL, CNRS, Sorbonne Universit\'e, Universit\'e de Paris, 24 rue Lhomond, 75005 Paris, France}
\author{C. Voisin}
\affiliation{Laboratoire de Physique de l'Ecole normale sup\'erieure, ENS, Universit\'e
PSL, CNRS, Sorbonne Universit\'e, Universit\'e de Paris, 24 rue Lhomond, 75005 Paris, France}
\author{B. Pla\c{c}ais} \email{bernard.placais@phys.ens.fr}
\affiliation{Laboratoire de Physique de l'Ecole normale sup\'erieure, ENS, Universit\'e
PSL, CNRS, Sorbonne Universit\'e, Universit\'e de Paris, 24 rue Lhomond, 75005 Paris, France}
\author{E. Baudin} \email{emmanuel.baudin@phys.ens.fr}
\affiliation{Laboratoire de Physique de l'Ecole normale sup\'erieure, ENS, Universit\'e
PSL, CNRS, Sorbonne Universit\'e, Universit\'e de Paris, 24 rue Lhomond, 75005 Paris, France}

\begin{abstract}
This Supplementary Information consists of a Table summarizing the  geometrical and electrical parameters of the ten hBN-encapsulated graphene transistors series. It is augmented by three figures detailing the DC transport (Fig.\ref{Fig1}), velocity flicker (Fig.\ref{Fig2}), and noise thermometry (Fig.\ref{Fig3}) of $6$ representative transistors discussed in the main text.
\end{abstract}

\maketitle

\renewcommand{\thefigure}{SI-\arabic{figure}}  
\renewcommand{\thetable}{SI-\arabic{table}}

\begin{table}[t]
\centering
\hspace*{-0.7cm}
    \begin{tabular}{| p{2.0cm} |  c |  c | c | c | c | c | c |  c |  c |  c |  c |  c |c|c|}
    \hline
    \textbf{Sample} & \textbf{Gate} & \textbf{hBN} & \textbf{L} & \textbf{W}& \boldmath{$t_{hBN}$} & \boldmath{$R_{c}$} & \boldmath{$\;\mu(0)\;$}&\boldmath{$\;v_{sat}\;$} & \boldmath{$\sigma_{Z}$} & \large{\boldmath{$A$}}& \boldmath{$E_{\Lambda}$} & \boldmath{$\mathrm{dP/dT_{N}}$}\\
       name & &grade & $\;\mu m\;$ &
    $\;\;\mu m\;\;$& $nm$ & $\;\;\;\Omega\;\;\;$ & $\;\mathrm{m^{2}/Vs }   $ &$\;10^6\;\mathrm{m/s}$ &  $\;\; \mathrm{mS}\;\;$ &$\;nm^{2}\;$ &$\;V/mm$ &$\mathrm{MW/Km^{2}}$  \\ \hline\hline
    \textbf{Lyon1}& Au & PDC & 9 & 9 & 142 & 250 & \textbf{4} & 0.82 & 0.1 & 0.64 & 67 & 0.068 \\ \hline
    \textbf{Lyon2}& Au & PDC & 5 & 5 & 142 & 80 & \textbf{8.5}  & 0.60 & 0.45& 0.54 & 99 & 0.402 \\ \hline
    \textbf{InOut1}& Au & HPHT & 12.5 & 17.5 & 164 & 71  & \textbf{6} & 0.33 & 0.40 & 0.57 & 48 & 0.373 \\ \hline
    \textbf{InOut2}& Au & HPHT & 20 & 8.5 & 118 & 300  & \textbf{11} & 0.46 & 0.55 & 0.22 & 94 & 0.128 \\ \hline
    \textbf{GrS5}& Graphite & HPHT & 3.8 & 4 & 98 & 100 &\textbf{8}  & 0.48 & 0.60 & 0.36 & 56 & 0.461 \\ \hline
    \textbf{AuS3}& Au & HPHT & 11.1 & 11.4 & 90 & 95 & \textbf{15} & 0.42 & 0.40 & 0.54 & 57 & 0.258 \\ \hline
    \textbf{Goyave2}& Au & HPHT & 9 & 13 & 84 & 89  & \textbf{2} & 0.22 & 0.40 & 0.23 & 86 & 0.585 \\ \hline
    \textbf{Flicker162}& Au & HPHT & 7.1 & 7.5 & 150 & 95 & \textbf{14.5} & 0.58 & 1.05 & 0.18 & 81 & 0.952 \\ \hline
    \textbf{Largevin2}& Si/SiO$_{2}$ & HPHT & 5.4 & 25.4 & 59 &40  & \textbf{6} & 0.36 & 0.25 & 0.61 & 89 & 0.452 \\ \hline
    \textbf{Coolox}& Si/AlOx & HPHT & 5.3& 5.8& 86& 140& \textbf{2}& 0.38& 0.27& 0.26 & 140 & 0.262 \\ \hline
    \end{tabular}
\caption{Geometrical and electrical properties of the $10$ devices series. The first 6 devices are analyzed in Fig.3 of the main text.  Basic properties include the nature of the gate electrode (Au or Graphite bottom gating, or back-gating with Si/SiO$_{2}$ or Si/AlOx) and the origin of the hBN dielectric, the channel length $L$, width $W$,  hBN dielectric thickness $t_{hBN}$, and contact resistance $R_{c}$. We have used two hBN grades: the high-pressure high-temperature (HPHT) from NIMS  and the polymer-derived ceramic (PDC) from Lyon. The mobility $\mu(0)$,  saturation velocity $v_{sat}$ and Zener interband conductivity $\sigma_{Z}$ are extracted from DC measurements using Eq.(1) of the main text. The next two columns contain the two flicker parameters $A$ and $E_\Lambda$ deduced from the two-fluid flicker analysis (see Fig.3 of the main text), which are discussed in the main text. The last column contains the values of the thermal resistance to the hBN substrate $\mathrm{dT_{N}/dP}$ in the Zener interband regime, extracted from thermal noise measurements.}
\label{tab:label}
\end{table}

\begin{figure}[h!]
\hspace*{-1cm}
\centering{}\includegraphics[width=18cm]{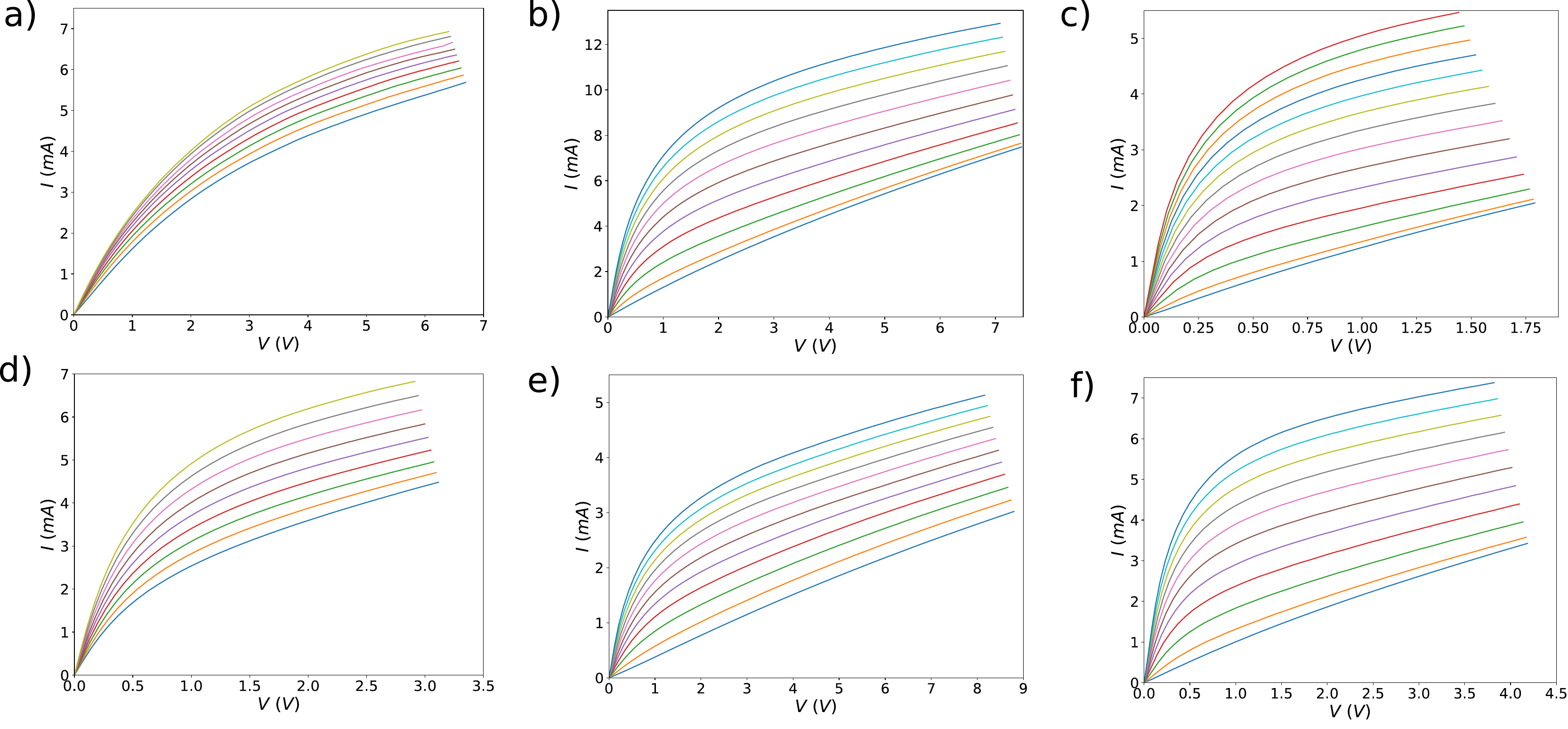} \caption{Room temperature current voltage $I(V)$ characteristics of $6$ typical high-mobility graphene transistors :  Lyon1 (a),  InOut1 (b), GrS5 (c), Lyon2 (d), InOut2 (e), AuS3 (f). Their geometrical and electrical parameters are described in Table \ref{tab:label}. The voltage drop across the contact resistance has been subtracted to access the local electric field $E=V/L$ where $L$ is the channel length. Main differences lie in  the mobility ranging from $\mu=4\;\mathrm{m^2/Vs}$ for Lyon1 (panel a) to $\mu=15\;\mathrm{m^2/Vs}$ for AuS3 (panel f), and the Zener conductivity ranging from $\sigma_Z=0.1\;\mathrm{mS}$ for Lyon1 (panel a) to $\sigma_Z=0.5\;\mathrm{mS}$ for GrS5 (panel c).}
\label{Fig1}
\end{figure}

\begin{figure}[h!]
\hspace*{-1cm}
\centering{}\includegraphics[width=18cm]{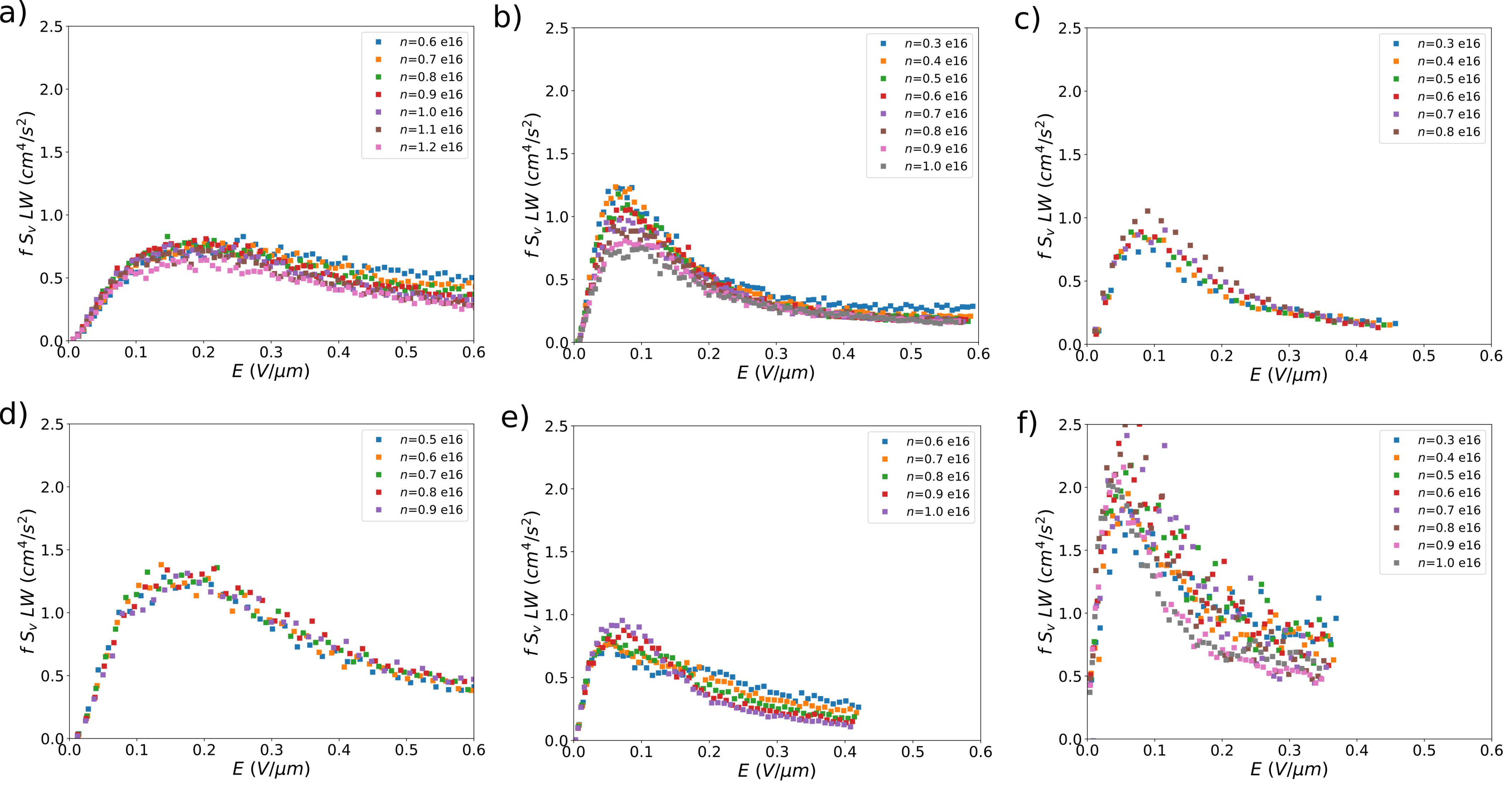} \caption{Intensive velocity flicker noise $f S_vLW(E)$ as function of electric field $E=V/L$ for $6$ typical high-mobility graphene transistors :  Lyon1 (a),  InOut1 (b), GrS5 (c), Lyon2 (d), InOut2 (e), AuS3 (f). Their geometrical and electrical parameters are described in Table \ref{tab:label}. Velocity noise $S_v=S_I/n^2e^2W^2$ is deduced from the measured current noise $S_I$ accounting for electrostatic doping $n$ and sample width $W$. The figure illustrates the scaling property of the velocity flicker as discussed in the main text. Sample statistics reveal a  rather universal velocity-flicker amplitude $f S_vLW=1$-$2\,\mathrm{cm^4/s^2}$, and significant variability in the electric-field dependence.}
\label{Fig2}
\end{figure}

\begin{figure}[h!]
\hspace*{-1cm}
\centering{}\includegraphics[width=18cm]{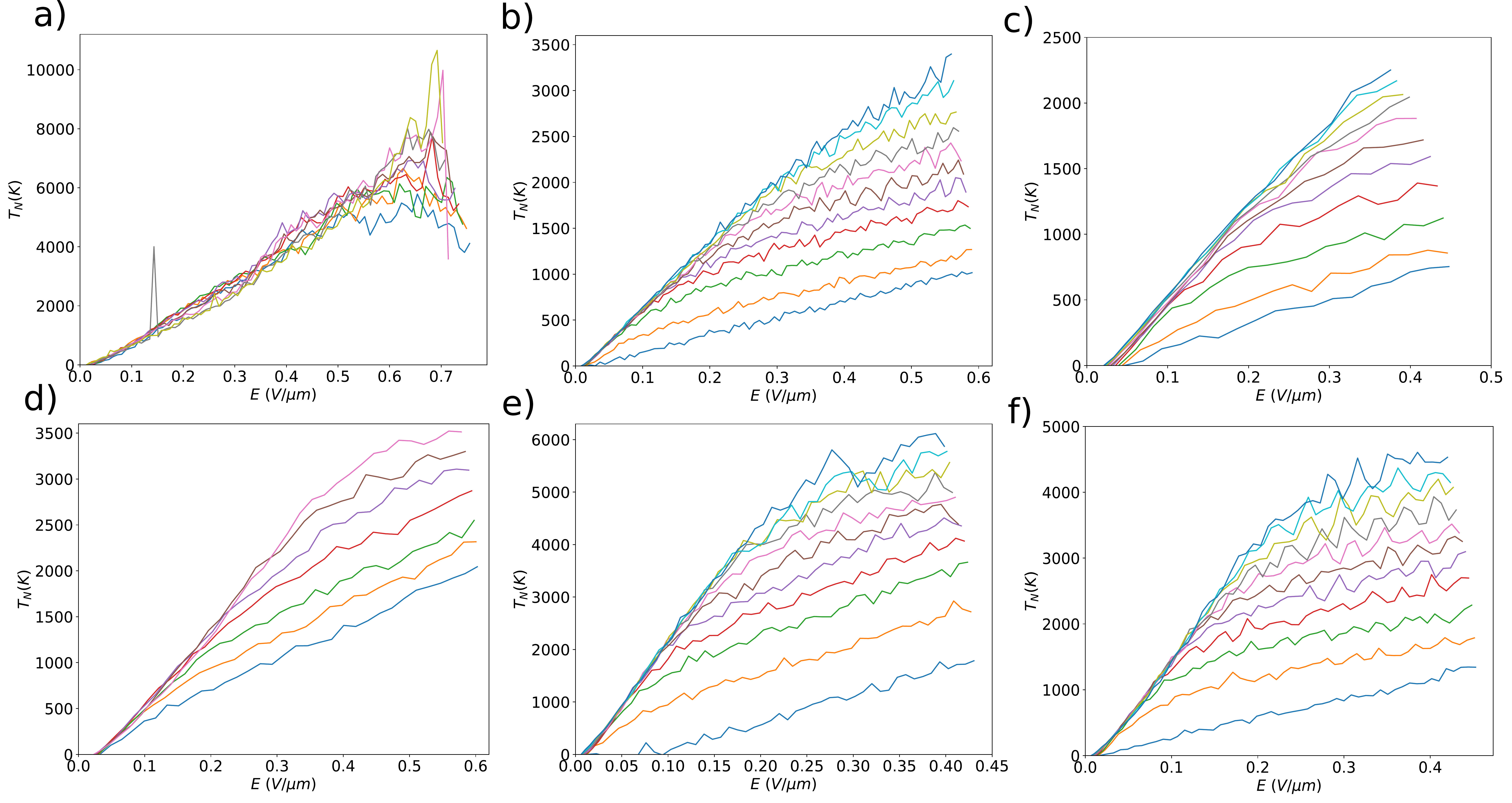} \caption{Noise temperature $T_N(E)$ as function of electric field $E$ for $6$ typical high-mobility graphene transistors :  Lyon1 (a),  InOut1 (b), GrS5 (c), Lyon2 (d), InOut2 (e), AuS3 (f). Their geometrical and electrical parameters are described in Table \ref{tab:label}. $T_N(E)$ is deduced from the Johnson-Nyquist noise $S_I=4Gk_BT_N$, measured in the flicker-free $1$-$10\;\mathrm{GHz}$ microwave band, taking the DC differential conductance $G$ as a prefactor. It characterizes the cooling pathways at stake in increasing $E$: the in-plane heat conduction at low $E$ which depends on mobility, and the radiative cooling by hyperbolic phonon polariton (HPhP) emission in the hBN substrate in the Zener regime at large $E$. Significant sample to sample variability is observed that reflects the sensitivity of hot-electron temperature to electronic mobility at low-$E$ and to the mobility of hBN-HPhPs at large-$E$ with a reduced HPhP cooling in the hBN-PDC-grade transistors Lyon1 (panel a) and Lyon2 (panel d), with respect to the hBN-HPHP-grade transistors (other panels).  }
\label{Fig3}
\end{figure}